\documentclass[12pt,journal,final,letterpaper,onecolumn,oneside]{IEEEtran}

\usepackage{cite,verbatim,balance}
\usepackage[cmex10]{amsmath}
\usepackage{amsfonts,algorithmic,array,graphicx}

\def\rR{{\mathbb R}}

\begin{document}
\title{A note on the sum-rate-distortion function of some lossy source
  coding problems involving infinite-valued distortion functions}

\author{Prakash~Ishwar \\ Department
of Electrical and Computer Engineering \\ Boston University, Boston,
MA 02215 USA \\ e-mail: {\tt pi@bu.edu}.}%

\maketitle

\begin{abstract}
For a number of lossy source coding problems it is shown that even if
the usual single-letter sum-rate-distortion expressions may become
invalid for non-infinite distortion functions, they can be approached,
to any desired accuracy, via the usual valid expressions for
appropriately truncated finite versions of the distortion functions.
\end{abstract}

\begin{IEEEkeywords}
Source coding, rate-distortion function, sum-rate, non-finite
distortion function.
\end{IEEEkeywords}

\section{Introduction}\label{sec:intro}
In a number lossy source coding problems, the minimum sum-rate needed
to attain a target expected distortion of $D$ or less -- the
operational sum-rate-distortion function -- is characterized by an
information-theoretic rate-distortion function $R(D)$ which is
expressed in terms of an optimization problem of the following form:
\begin{equation}
R(D) := \inf_{p_{X,Y,U,\widehat{X}} \in \mathcal{A} \mbox{ s.t. }
  E[d(X,Y,U,\widehat{X})] \leq D} f(p_{X,Y,U,\widehat{X}}),
\label{eqn:RD}
\end{equation}
where $X \in \mathcal{X}, Y \in \mathcal{Y}, U \in \mathcal{U},$ and
$\widehat{X} \in \widehat{\mathcal{X}}$ are, respectively, the source,
side-information, auxiliary, and reconstruction random variables
taking values in finite alphabets, $p_{X,Y,U,\widehat{X}}$ is their
joint pmf, $f$ is a finite linear combination of conditional mutual
informations involving some or all the random variables, $\mathcal{A}$
is a finite set of marginal consistency\footnote{Specifically,
  $\sum_{u,\widehat{x}} p_{X,Y,U,\widehat{X}}(x,y,u,\widehat{x}) =
  p_{X,Y}(x,y)$ for all $x,y$.} and Markov-chain constraints that the
random variables need to satisfy, and $d$ is a real-valued,
nonnegative distortion function which does not depend on $U$. Examples
include Shannon rate-distortion \cite{Shannon59} and Gray-Leiner
conditional rate-distortion \cite{Gray73,Leiner73,Gray-Leiner74} where
auxiliary random variables are not needed, Wyner-Ziv rate-distortion
\cite{Wyner-Ziv76,Wyner78} which uses a single auxiliary random
variable, and Kaspi's two-way rate-distortion \cite{Kaspi85} which
uses multiple auxiliary random variables (collectively denoted by $U$
here for convenience).

For Shannon's lossy source coding problem, the fact that the usual
information-theoretic expression of the form (\ref{eqn:RD}) will
continue to coincide with the operational rate-distortion function
when the distortion function $d$ can take the value $\infty$, was
established by Pinkston in \cite{Pinkston67} (also see
\cite{Gallager68}[Ch~9,Historical Notes and References]). That this
should also be true for the Gray-Leiner conditional rate-distortion
function should be expected, since the encoder can group source
samples that have the same side-information value into multiple
conditional sources and code them separately like in Shannon's lossy
source coding problem. We have, however, been unable to locate a
reference which discusses the extension of the Gray-Leiner conditional
rate-distortion function to non-finite distortion functions.

An example of an infinite-valued distortion function is the so-called
erasure distortion
function~\cite{Cover-Thomas91}[Chapter~13, Problem~7].  Here, the
distortion function equals zero if the source and reconstruction
symbols agree, it equals one if the reconstruction is a special
erasure symbol (irrespective of the source symbol), and it equals
infinity otherwise. The erasure distortion function has appeared in a
number of recent works: in \cite{Wagner-Ananthram08}[Sec.~III-B] in
the context of the CEO problem, in \cite{Ahmed-Wagner12} in the
context of the multiple descriptions coding problem, and in
\cite{Courtade-Wesel11} in the context of some examples.

The erasure distortion function was used in \cite{Ma-Ishwar10} and
\cite{Ma-Ishwar13}[Secs.~VII.B,C] to construct the first example which
shows that the usual information-theoretic two-way rate-distortion
function with two messages can be strictly smaller than the usual
one-message Wyner-Ziv information-theoretic rate-distortion
function. Specifically, it was shown that for sufficiently correlated
doubly-symmetric binary sources \cite{Wyner-Ziv76}[Sec.II,Eqn.(20)]
and the erasure distortion function, it is possible to make the ratio
of the one-message rate to the two-message sum-rate arbitrarily large
while simultaneously making the ratio of the backward rate to the
forward rate in the two-message sum-rate arbitrarily small.

However, in~\cite{Chia-Chong14} it was shown that, contrary to the
claim made in footnote 8 of \cite{Ma-Ishwar13}[Appendix E], the
{usual} information-theoretic rate-distortion functions for the
Wyner-Ziv and two-way source coding problems, which are used in
\cite{Ma-Ishwar10} and \cite{Ma-Ishwar13}[Secs.~VII.B,C], are in fact
\emph{strictly smaller} than their operational counterparts for the
erasure distortion function and doubly-symmetric binary sources which
satisfy a positivity condition. This implies that the usual
information-theoretic rate-distortion expressions are not always valid
for non-finite distortion functions, such as the erasure distortion
function, in a distributed source coding setting even though they are
for the Shannon and Gray-Leiner settings.

For the Wyner-Ziv and two-way source coding problems, the {correct}
information-theoretic rate-distortion expressions that coincide with
their operational counterparts were characterized
in~\cite{Chia-Chong14} for the erasure distortion function when the
source and side information satisfy a positivity condition. It was
shown, in particular, that the two-message two-way rate-distortion
function for this problem exactly coincides with the one-message
Wyner-Ziv rate-distortion function casting into doubt some of the
conclusions reached in \cite{Ma-Ishwar10} and
\cite{Ma-Ishwar13}[Secs.~VII.B,C] about the benefit of interaction
(two-way coding) for lossy source reproduction.

The aim of this article is to demonstrate that all the conclusions
reached in \cite{Ma-Ishwar10} and \cite{Ma-Ishwar13}[Secs.~VII.B,C]
regarding the benefit of interaction for lossy source reproduction are
correct if one amends footnote 8 of \cite{Ma-Ishwar13}[Appendix E] to
the following: ``Although the usual information-theoretic expressions
for the Wyner-Ziv and two-message rate-distortion functions are not
operationally attainable for the non-finite erasure distortion
function, they can be operationally approached, as closely as desired,
by replacing the $\infty$ value in the erasure distortion function
with a sufficiently large, but finite, positive real number.''. The
possibility of making such an approach work is contained in a
suggestion of anonymous reviewers reported in Remark~1.1 of
\cite{Chia-Chong14}. Such an approach is also taken in
\cite{Wagner-Ananthram08}[Sec.~III-B] where it is remarked that an
infinite-valued distortion measure is unforgiving of decoding errors
that have nonzero probability even if they are negligible. Decoding
errors with vanishingly small but nonzero probability are unavoidable
in most distributed source coding problems.

Formally speaking, our main result is that if $d_n$ is a sequence of
bounded distortion functions that monotonically increases to a
non-finite distortion function $d_{\infty}$, then the corresponding
sequence of information-theoretic rate-distortion functions $R_n(D)$
associated with $d_n$ also monotonically increases to the \emph{usual}
information-theoretic rate-distortion function $R_{\infty}(D)$
associated with $d_{\infty}$. Thus by making $d_n$ approach
$d_{\infty}$, it is possible to make the operational
sum-rate-distortion function for $d_n$ as close as desired to
$R_{\infty}(D)$ even though $R_{\infty}(D)$ itself may not be
operationally attainable for $d_{\infty}$. Thus intuitions, examples,
and broad qualitative conclusions (such as the benefit of interaction)
that can be formed on the basis of examining $R_{\infty}(D)$ will be
\emph{essentially} correct even from an operational perspective in the
sense that while they may not hold true operationally for $d_{\infty}$
itself, they will for $d_n$, for all sufficiently large $n$.

\section{Main result}

We first establish a general result and then discuss its application
to rate-distortion functions.

\subsection{Statement}\label{sec:statement}
Consider the following finite-dimensional optimization problem
\begin{equation}
\psi_n(D) := \inf_{p \in \mathcal{C}_n(D) := \mathcal{A} \cap
  \mathcal{B}_n(D)} f(p)
\label{eqn:psi-n}
\end{equation}
where $p$ is a probability vector (prob.vec.) in $\rR^k$, $k$ a finite
positive integer, $f$ is a real-valued continuous function of $p$,
$\mathcal{A}$ is a fixed, nonempty, compact subset of probability
vectors in $\rR^k$, $D$ is a finite nonnegative real number, and
$\mathcal{B}_n(D) := \{p \mbox{ prob.vec.}: \langle p, d_n \rangle
\leq D\}$, where $\langle \cdot, \cdot \rangle$ denotes the usual
Euclidean-space inner product and $d_n$, $n = 1, 2, \ldots$, is a
sequence of vectors in $\rR^k$ with finite, nonnegative, and
nondecreasing components some of which (but not all) increase to
$\infty$ while the rest monotonically increase to some finite
nonnegative real numbers. We denote the limit of the $d_n$'s by
$d_{\infty}$. Thus, $\mathcal{B}_n(D)$ (and therefore also
$\mathcal{C}_n(D)$) is a nested nonincreasing sequence of compact
subsets of probability vectors in $\rR^k$. We define
\begin{equation}
\mathcal{B}_{\infty}(D) := \cap_{n=1}^{\infty} \mathcal{B}_n(D) = \{p
\mbox{ prob.vec.}: \langle p, d_{\infty} \rangle \leq D\}
\label{eqn:Binf}
\end{equation}
with the convention $0 \cdot \infty := 0$ in the inner product
and observe that the support of any $p$ in $\mathcal{B}_{\infty}(D)$
excludes components where $d_{\infty}$ equals $\infty$ (since $D$ is
finite) and that $\mathcal{B}_{\infty}$ is a compact subset of
probability vectors in $\rR^k$.
Let 
\begin{equation}
\psi_{\infty}(D) := \inf_{p \in \mathcal{C}_{\infty}(D) := \mathcal{A}
  \cap \mathcal{B}_{\infty}(D)} f(p).
\label{eqn:psi-inf}
\end{equation}
The main result is that $\psi_n(D) \uparrow \psi_{\infty}(D)$ for all
finite nonnegative $D$ for which $\mathcal{C}_{\infty}(D)$ is
nonempty. 

\subsection{Proof}
To prove this result, first observe that the constraint sets
$\mathcal{C}_n(D)$ are nested, nonincreasing, and contain
$\mathcal{C}_{\infty}(D)$. This implies that
\[
\psi_n(D) \uparrow \lim_{n\rightarrow \infty}\psi_n(D) \leq \psi_{\infty}(D).
\]
The main result is proved by establishing the reverse
inequality. Towards this end, we note that the minimands in
(\ref{eqn:psi-n}) and (\ref{eqn:psi-inf}) are continuous functions and
that the constraint sets are compact. This implies that there is a
sequence of probability vectors $p_n^{(D)} \in \mathcal{C}_n(D)$ and a
probability vector $p_{\infty}^{(D)} \in \mathcal{C}_{\infty}(D)$ such
that $\psi_n(D) = f(p_n^{(D)})$ for all $n$ and $\psi_{\infty}(D) =
f(p_{\infty}^{(D)})$. Since all these probability vectors belong to
the compact set $\mathcal{C}_1(D)$, there is a subsequence
$p_{n_j}^{(D)}, j = 1, 2, \ldots$, converging to a probability vector
$q_{\infty}^{(D)}$. We will shortly show that $q_{\infty}^{(D)}$ is in
$\mathcal{C}_{\infty}(D)$. Then, since $f$ is continuous,
$\psi_{n_j}(d) = f(p_{n_j}^{(d)}) \uparrow f(q_{\infty}^{(D)}) \geq
\psi_{\infty}(D)$ establishing the reverse inequality. Finally, to see
why $q_{\infty}^{(D)}$ is in $\mathcal{C}_{\infty}(D)$, note that for
components of $d_{n_j}$ that increase to infinity, the corresponding
components of $p_{n_j}^{(D)}$ must converge to zero since $D$ is
finite. The remaining components of both $d_{n_j}$ and $p_{n_j}^{(D)}$
converge to finite nonnegative real numbers. Thus the subsequence of
their (nonnegative) inner products, which is no more than $D$ (finite and
nonnegative), converges to the inner product of their limits.

\subsection{Application to rate-distortion functions}

Comparing (\ref{eqn:RD}) and (\ref{eqn:psi-n}) it is apparent that
they have the same form. To demonstrate that (\ref{eqn:RD}) is, in
fact, a special case of (\ref{eqn:psi-n}), we only need to verify that
the minimand, minimizing variable, and constraint sets of
(\ref{eqn:RD}) satisfy all the assumptions in Sec.~\ref{sec:statement}
that (\ref{eqn:psi-n}) is required to satisfy. First note that $p$ in
(\ref{eqn:psi-n}) corresponds to $p_{X,Y,U,\widehat{X}}$ in
(\ref{eqn:RD}) with $k = \mid \mathcal{X} \times \mathcal{Y} \times
\mathcal{U} \times \widehat{\mathcal{X}} \mid$. The $f$ in
(\ref{eqn:RD}) is a finite linear combination of conditional mutual
informations involving some or all the variables
$X,Y,U,\widehat{X}$. This is a real-valued (in fact also nonnegative
and bounded) continuous function of $p_{X,Y,U,\widehat{X}}$ since
conditional mutual informations are continuous functions of the joint
pmf of all the variables that appear in them
\cite{Yeung02}[Ch.2,Sec.2.3], marginal pmfs are linear (therefore
continuous) functions of the joint pmf, and the composition of a
finite number of continuous functions is continuous. The set
$\mathcal{A}$ in (\ref{eqn:RD}) is the set of pmfs which satisfy
certain marginal consistency and Markov-chain constraints associated
with the random variables $X, Y, U, \widehat{X}$. This set is compact
because a marginal consistency constraint is a linear equality
constraint on the joint pmf (hence it defines a closed hyperplane
within the bounded simplex of joint pmfs) and Markov-chain constraints
can be expressed as the zero-level sets of appropriate conditional
mutual information functions which, as we just discussed, are
continuous. The distortion function $d$ in (\ref{eqn:RD}) corresponds
to $d_n$ in (\ref{eqn:psi-n}). They are both real-valued (finite) and
nonnegative. The set $\mathcal{B}_n(D)$ in (\ref{eqn:psi-n}) then
corresponds to the expected distortion constraint
$E[d(X,Y,U,\widehat{X})] \leq D$ in (\ref{eqn:RD}). Finally, in
typical scenarios of the source coding problems discussed in
Sec.~\ref{sec:intro} (including \cite{Ma-Ishwar10} and
\cite{Ma-Ishwar13}[Secs.~VII.B,C]), $D$ is finite, and the feasible
set $\{p_{X,Y,U,\widehat{X}} \in \mathcal{A} \mbox{ s.t. }
E[d(X,Y,U,\widehat{X})] \leq D\}$ is nonempty.

\section*{Acknowledgment} The author would like to thank Nan Ma,
Sandeep Pradhan, and Bob Gray for their feedback on an earlier draft
of this work.


\begin{thebibliography}{1}

\bibitem{Shannon59}
%
C.~E.~Shannon, ``Coding theorems for a discrete source with a fidelity
criterion,'' in \emph{IRE Conv.~Rec.}, vol.~7, 1959,
pp.~142--163. (Also in \emph{Information and Decision Processes},
R.~E.~Machol, Ed.~New York: McGraw-Hill, 1960, pp. 93--126, and in
\emph{Claude Elwood Shannon: Collected Papers}, N.~J.~A.~Sloane and
A.~D.~Wyner, Eds.~Piscataway, NJ: IEEE Press, 1993, pp.~325-350.)

\bibitem{Gray73}
%
R.~M.~Gray, ``A new class of lower bounds to information rates of
stationary sources via conditional rate-distortion functions,''
\emph{IEEE Trans.~Inform.~Theory}, vol.~IT-19, pp.~480--489, 1973.

\bibitem{Leiner73} 
%
B.~M.~Leiner, ``Rate-distortion theory for sources with side
information,'' Ph.~D.~dissertation, Stanford Univ., Stanford, CA,
Aug.~1973.

\bibitem{Gray-Leiner74}
%
B.~M.~Leiner and R.~M.~Gray, ``Rate-distortion for ergodic sources
with side information,'' \emph{IEEE Trans.~Inform.~Theory},
vol.~IT-20, pp.~672--675, 1974.

\bibitem{Wyner-Ziv76}
%
A.~D.~Wyner and J.~Ziv, ``The rate-distortion function for source
coding with side-information at the receiver,'' \emph{IEEE
  Trans.~Inform.~Theory}, vol.~IT-22, pp.~1--11, 1976.

\bibitem{Wyner78}
%
A.~D.~Wyner, ``The rate-distortion function for source coding with
side information at the decoder-II: General sources,''
\emph{Inform.~Contr.}, vol.~38, pp.~60--80, 1978.

\bibitem{Kaspi85}
%
A.~H.~Kaspi, ``Two-way source coding with a fidelity criterion,''
\emph{IEEE Trans.~Inform.~Theory}, vol.~IT-31, pp.~735--740, 1985.

\bibitem{Pinkston67}
%
J. ~T.~Pinkston, ``Encoding Independent Sample Information Sources,''
\emph{MIT Research Laboratory of Electronics, Technical Report 462},
1967.

\bibitem{Gallager68}
%
R.~G.~Gallager, \emph{Information Theory and Reliable Communication},
John Wiles \& Sons, Inc., 1968.

\bibitem{Cover-Thomas91}
%
T.~M.~Cover and J.~A.~Thomas, \emph{Elements of Information Theory},
John Wiles \& Sons, Inc., 1991.

\bibitem{Wagner-Ananthram08}
%
A.~B.~Wagner and V.~Ananthram, ``An improved outer bound for multi
terminal source coding,'' \emph{IEEE Trans.~Inform.~Theory},
vol.~IT-54, pp.~1919--1937, 2008.

\bibitem{Courtade-Wesel11}
%
T.~Courtade and R.~Wesel, ``Multiterminal source coding with an
entropy-based distortion measure,'' in \emph{Proc.~IEEE International
  Symposium on Information Theory}, St.~Petersburg, Russia, Jul.~31 --
Aug.~5, 2011, pp.~2040--2044.

\bibitem{Ahmed-Wagner12}
%
E.~Ahmed and A.~B.~Wagner, ``Erasure multiple descriptions,'' \emph{IEEE
  Trans.~Inform.~Theory}, vol.~IT-58, pp.~1328--1344, 2012.

\bibitem{Ma-Ishwar10}
%
N.~Ma and P.~Ishwar, ``Interaction strictly improves the Wyner-Ziv
rate-distortion function,'' in \emph{Proc.~IEEE International
  Symposium on Information Theory}, Austin, TX, USA, Jun.~13-18, 2010,
pp.~61--65.

\bibitem{Ma-Ishwar13}
%
N.~Ma and P.~Ishwar, ``The infinite-message limit of two-terminal
interactive source coding,'' \emph{IEEE Trans.~Inform.~Theory},
  vol.~IT-59, pp.~4071--4094, 2013.

\bibitem{Chia-Chong14}
%
Y-K.~Chia and H.~F.~Chong, ``On lossy source coding with side
information under the erasure distortion measure,'' in
\emph{Proc.~IEEE International Symposium on Information Theory},
Honolulu, HI, USA Jun.~29 -- Jul.~4, 2014.

\bibitem{Yeung02}
%
R.~W.~Yeung, \emph{A First Course in Information Theory}, Kluwer
Academic~/~Plenum Publishers, New York, NY, 2002.

\end{thebibliography}
\end{document}